\documentclass[preprint,aps,pre,amsmath,superscriptaddress]{revtex4-1}
\usepackage[utf8]{inputenc}
\usepackage[T1]{fontenc}
\usepackage{newtxtext}
\usepackage[subscriptcorrection,vvarbb,varg]{newtxmath}
\usepackage{graphicx}
\usepackage{hyperref}
\hypersetup{colorlinks=true,breaklinks=true,allcolors=blue}
\usepackage{siunitx,xcolor}
\usepackage[normalem]{ulem}
\definecolor{cc}{rgb}{0,0,0}

\makeatletter
\def\fps@figure{htb}
\makeatother

\begin{document}
\title{Characterization of SOL Plasma Flows and Potentials in ICRF-Heated Plasmas in Alcator C-Mod}
\author{R. Hong}
\affiliation{Center for Energy Research, University of California San Diego, La Jolla, CA 92093, USA}
\author{S. J. Wukitch}
\author{Y. Lin}
\author{J. L. Terry}
\affiliation{MIT Plasma Science and Fusion Center, Cambridge, Massachusetts 02139, USA}
\author{I. Cziegler}
\affiliation{York Plasma Institute, Department of Physics, University of York, Heslington, York, YO10 5DD, UK}
\author{M. L. Reinke}
\affiliation{Oak Ridge National Laboratory, Oak Ridge, Tennessee 37831, USA}
\author{G. R. Tynan}
\affiliation{Center for Energy Research, University of California San Diego, La Jolla, CA 92093, USA}
\affiliation{Center for Fusion Sciences, Southwestern Institute of Physics, Chengdu, Sichuan 610041, China}

\date{\today}

\begin{abstract}
  \normalsize
  Gas-puff imaging techniques are employed to determine the far SOL region radial electric field and the plasma potential in ICRF heated discharges in the Alcator C-Mod tokamak.
  The 2-dimensional velocity fields of the turbulent structures, which are advected by RF-induced $ \mathbf{E\times B} $ flows, are obtained via the time-delay estimation (TDE) techniques.
  Both the magnitude and radial extension of the radial electric field $ E_r $
  are observed to increase with the toroidal magnetic field strength $ B_\varphi $ and the ICRF power.
  In particular, the RF-induced $ E_r $ extends from the vicinity of the ICRF antenna to the separatrix when $ B_\varphi=7.9\,\mathrm{T} $ and $ P_{\mathrm{ICRF}} \gtrsim 1\,\mathrm{MW} $.  
  In addition, low-Z impurity seeding near the antenna is found to substantially reduce the sheath potential associated with ICRF power.
  The TDE techniques have also been used to revisit and estimate ICRF-induced potentials in different antenna configurations: (1) conventional toroidally-aligned (TA) antenna versus field-aligned (FA) antenna; (2) FA monopole versus FA dipole.
  It shows that FA and TA antennas produce similar magnitude of plasma potentials, and the FA monopole induced greater potential than the FA dipole phasing.
  The TDE estimations of RF-induced plasma potentials are consistent with previous results based on the poloidal phase velocity.
\end{abstract}

\maketitle

\section{Introduction}

Ion cyclotron range of frequency (ICRF) heating is one of a few promising auxiliary heating techniques for achieving thermonuclear fusion relevant temperatures in magnetic confinement fusion devices.
Furthermore, 20 MW of ICRF power is planned for in ITER for plasma heating and current drive \cite{Swain2007FEaD603}.
While demonstrated to efficiently heat D-T plasmas to thermonuclear temperatures, e.g. TFTR \cite{Wilson1995PRL842} and JET \cite{Start1998PRL4681}, ICRF heating is often associated with enhanced core impurity contamination, which makes it incompatible with high performance plasmas \cite{Lipschultz2001NF585,Wukitch2013PoP,Bobkov2016NF84001}.
Although the mechanism is not fully understood, the RF-enhanced plasma potential associated with ICRF power has been thought to play a significant role through increased sputtering near the ICRF antennas \cite{Wukitch2013PoP,Garrett2012FEaD1570}.
RF-enhanced plasma potentials and electric fields have been observed in the vicinity of the
energized RF antennas in C-Mod and Tore-Supra  \cite{Faudot2010PoP,Garrett2012FEaD1570,Myra2008PRL195004,Myra2010PPaCF15003,Cziegler2012PPaCF105019,Ochoukov2014PPaCF15004}.

The basic physical picture for RF-enhanced plasma potentials is well known \cite{Butler1963PoF1346,Nieuwenhove1989JoNM288,Perkins1989NF583}.
Electrons respond faster to the parallel electric field of RF waves ($ E_\parallel $) than ions, and develop rectified sheath potentials to balance the electron and ion fluxes over a RF-cycle and maintain time-averaged quasi-neutrality.
This rectified sheath potential is associated with the slow wave (SW) that can be generated directly from the antenna or be converted from unabsorbed fast waves (FWs) into SWs in the scrape-off layer (SOL) \cite{Myra2008PRL195004,Myra2010PPaCF15003}.
According to conventional models, the decay length of the SW is the same order as the plasma skin depth, $ \delta_\mathrm{pe} = c/\omega_\mathrm{pe} $, which is no more than a few millimeters in the far SOL region of tokamaks.
Some experimental observations, however, have confirmed that the radial expansion of RF-enhanced plasma potentials could be considerably larger than the skin depth \cite{Cziegler2012PPaCF105019,Ochoukov2014PPaCF15004}.
Since the discover of this `anomalous’ penetration depth of the dc potential structures, a number of models have been proposed to explain the underlying physics that broaden the radial structure of the RF-sheath \cite{Faudot2010PoP,Colas2012PoP92505,Colas2017PPaCF25014}.

In most experiments, the rectified sheath structures are directly measured by conventional or emissive Langmuir probes \cite{Ochoukov2014PPaCF15004,Ochoukov2013JoNM875}.
However, the operation of probes is often limited in RF-heated plasmas since the high heat flux easily shortens the lifetime of the probe. 
Moreover, probes can only measure the 1D profile of the RF sheath potential in each single shot, while 2D resolution would be highly useful as the currents in the ICRF antenna are expected to break the poloidal symmetry. 
Recently the gas-puffing-imaging (GPI) diagnostics have been utilized to detect the structures of the RF-enhanced plasma potentials in the Alcator C-Mod tokamak \cite{Cziegler2012PPaCF105019}.
The GPI diagnostics, being capable of observing a 2D area near the ICRF antenna in a single shot and at higher temperature, can overcome challenges of using probes in RF-heated plasmas.

In this study, we use the 2D GPI signals to estimate the velocity of the SOL turbulent structures, which are advected by the ICRF-induced $ \mathbf{E\times B} $ flows.
The 2D velocity field is calculated using the time-delay estimation (TDE) techniques.
Then the magnitude and radial extension of RF-enhanced electric fields and plasma potentials near the ICRF antennas can be determined.

\section{Experimental Setup}

\begin{figure}
\includegraphics[width=5in]{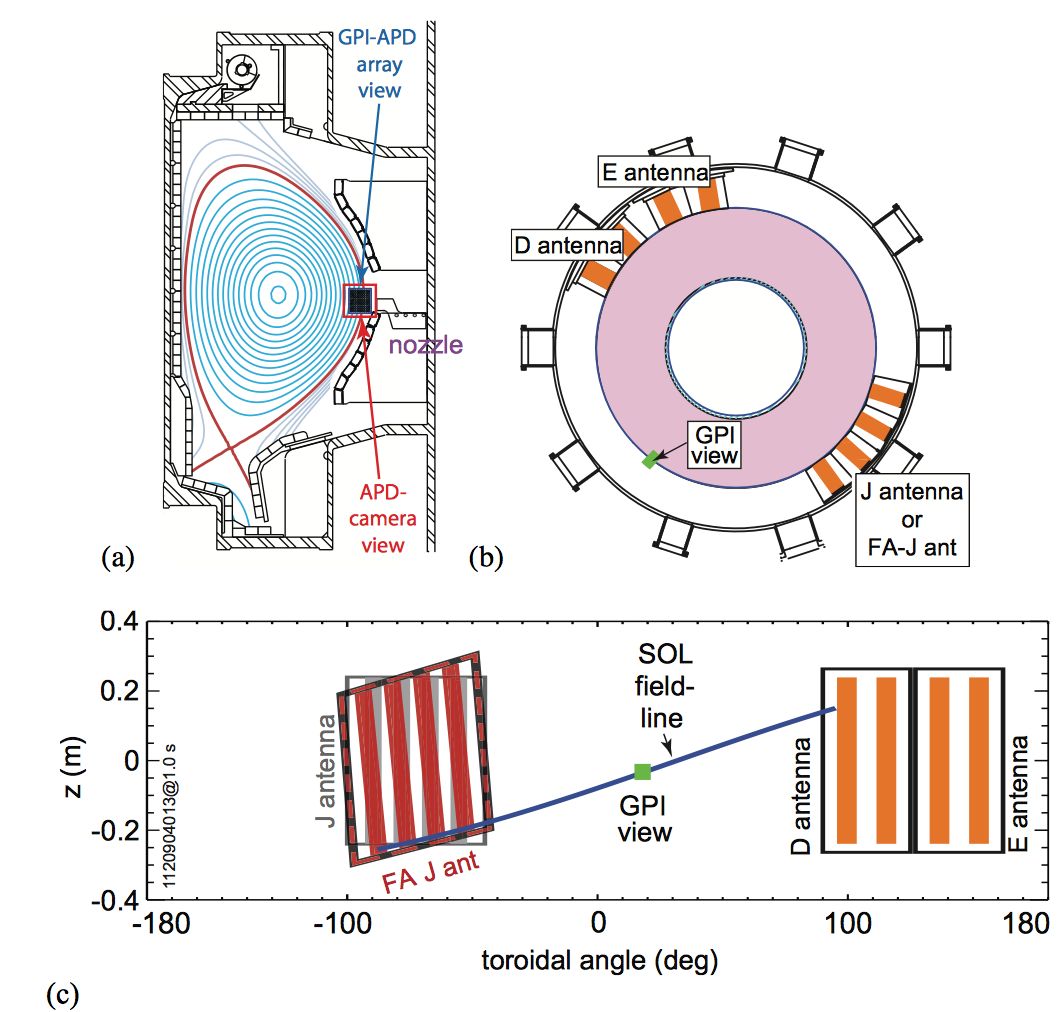}
\caption{\label{fig:cmod} (a) Poloidal X-section of C-Mod at the GPI toroidal location. GPI views are oversize for clarity. (b) Top view of C-Mod showing the GPI view in relation to the ICRF antennas. (c) The toroidal layout of the antennas and GPI view along with the field-line that maps from the GPI view to its magnetically connected location on the FA-J antenna. The ``TA-J antenna'' (in gray behind the ``FA-J ant'') was replaced by the “FA-J ant” in early 2012. Note that the straps of the ``TA-J antenna'' were vertical, while those of the ``FA-J ant'' are perpendicular to the local field-line for $ q_{95}=3.8 $.}
\end{figure}

The SOL plasma flows and ICRF-enhanced plasma potentials are measured in ICRF-heated deuterium majority plasmas in Alcator C-Mod tokamak in which the plasma-facing components (PFCs) are entirely composed of high-Z molybdenum tiles \cite{Hutchinson1994PoP11511}.
ICRF antennas are located at three different ports (Fig.~\ref{fig:cmod}(b)-(c)). 
More detailed descriptions of C-Mod and its ICRF antennas can be found in previous papers \cite{Hutchinson1994PoP11511,Greenwald2014PoP,Wukitch2013PoP,Bonoli2007FSaT401}.
All discharges in this study were operated in L-mode.
The plasma response to ICRF waves is diagnosed using the 2D gas-puffing imaging (GPI) system, which detects the line radiation of the injected neutral gas (He 586 nm in this study) and is able to trace the motion of the emissive turbulent structures \cite{Cziegler2010PoP,Cziegler2012PPaCF105019,Cziegler2013PoP}.
The diagnostic gas puffs are injected from a 4-barrel nozzle mounted on the low-field-side (LFS) limiter.
To avoid disturbing the measurements of RF-induced plasma flows, the amount of the helium gas used by GPI diagnostics is precisely controlled and is much less than that for impurity seeding.  Further GPI derived potentials were compared with emissive probe measurements and we found to agree \cite{Ochoukov2013JoNM875}.
In the discharges studied in this paper, the nozzle is 3 cm outside the separatrix and 2.54 cm below the midplane.
The field of view (FOV) of the 2D GPI array, as shown in Fig.~\ref{fig:cmod}(a), covers an area of $ 3.5\, \mathrm{cm} \times 3.9\,\mathrm{cm} $ in the radial-vertical plane. 
The imaging data are sampled at 2 MHz by avalanche photodiodes (APDs).
Although the FOV of GPI is toroidally separated from the energized ICRF antennas ($ L_\parallel \sim 2 \,\mathrm{m} $), it is magnetically connected with some parts of the D and J antenna, as can be seen in Fig.~\ref{fig:cmod}(c).
Therefore, the GPI system is able to measure the plasma potential on fields local to ICRF antennas. 
In 2012 the toroidally aligned (TA) J antenna was replaced by a ``field-aligned'' antenna, the ``FA-J ant'', whose four current straps and antenna box structure are perpendicular to the total magnetic field (for $ q_{95} \sim 3.8 $) \cite{Wukitch2013PoP}.
One of the primary motivations for installing the field-aligned antenna was to reduce integrated parallel RF electric fields, the very field whose sheath rectification is suspected to produce the RF-induced plasma potentials that are examined in this study.

\section{Plasma Flows and Electric Fields in the SOL Region}

\subsection{Time-Delay Estimation and 2D Velocity Field}

By using the data from the 2D APD GPI array, we are able to obtain the plasma flows induced by the rectified sheath potential in the SOL region via the time-delay estimation (TDE) technique.
In the TDE technique, the cross-correlations of the observed brightness fluctuations $ \tilde{I} $ from the 586 nm He I line emission on neighboring spots are calculated.
To estimate the time interval for fluctuations to propagate between two observation locations, the time lag of the maximum correlation, $ \tau_m $, can be obtained and the local velocities of the emissive structures are simply estimated as
\[ v_{ij} = \frac{d_{ij}}{\tau_m} ,\]
where $ d_{ij} $ is the distance between two observation points.
In this study, a time window of $ \tau \sim 0.1\, \mathrm{ms} $ is used for calculating the cross correlation, which corresponds to a sample length of 200 frames, introducing an effective Nyquist frequency of $ 10 \, \mathrm{kHz} $ in the velocity estimations.

\begin{figure*}
  \includegraphics[width=\textwidth]{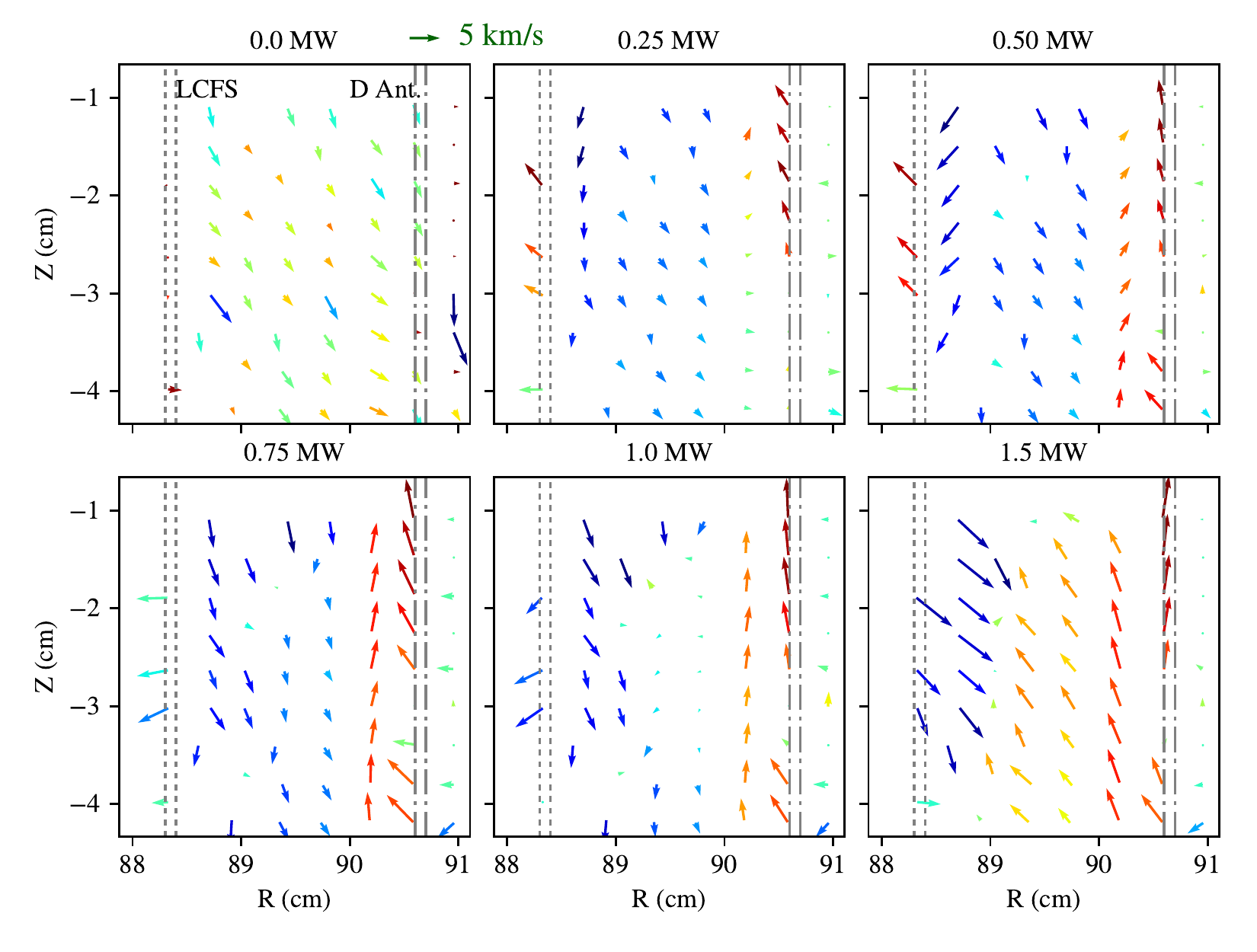}
  \caption{\label{fig:vel_field} Equilibrium velocity fields in the SOL region at different ICRF heating powers. Red arrows point in the EDD (upward); blue arrows point in the IDD (downward). Black dotted lines indicate the position of the LCFS; black dash-dot lines indicate the position of the D antenna. The green arrow at top of the figure indicates the scale of $ V = 5$ km/s. Clearly, the poloidal flows near antenna change direction as the ICRF power is increased.}
\end{figure*}

Typical equilibrium velocity fields of SOL plasmas at varied ICRF heating powers are shown in Fig.~\ref{fig:vel_field}.
Different colors stand for different directions and magnitudes of poloidal flows. 
Red arrows point in the electron diamagnetic direction (EDD) which is upward, while the blue arrows point in the ion diamagnetic direction (IDD) which is downward.
In ohmic plasmas, the poloidal flows in the SOL region is usually in the IDD direction, because the Bohm sheath drop leads to an outward radial electric field \cite{Stangeby2000}, $ E_r \approx - 3 \partial_r T_e/ e > 0 $, and thus a $ \mathbf{E\times B} $ poloidal velocity in the IDD, $ V_\theta \approx 3 \partial_r T_e/ eB_\varphi $, where $ B_\varphi $ is the strength of the toroidal magnetic field.
At low ICRF powers, poloidal flows near the LCFS are still in the IDD direction ($ 89 < R < 90 \,\si{cm} $), but poloidal velocities near the antenna change from IDD to EDD ($ 90 < R < \SI{91}{cm} $).
The changes of flow patterns near the antenna strongly suggest that large-scale flow patterns are induced on the field lines magnetically connected to active ICRF antennas.

In our previous study, radial profiles of the poloidal phase velocities were estimated from the experimental dispersion relation, i.e.~local conditional spectra $ S(k_\theta|f) $ \cite{Cziegler2012PPaCF105019}.
The TDE techniques used in this study have been benchmarked against the phase velocity estimations.
The velocity profiles from the TDE techniques are in agreement with previous estimations based on $ S(k_\theta|f) $ structures.
However, the poloidal phase velocity estimations are based on Fourier transform and simply assume that $ V_\theta $ is uniform along poloidal direction, while the TDE techniques are able to give more information such as 2D velocity fields and variations in $ V_\theta $ along the poloidal direction, as shown in Fig.~\ref{fig:vel_field}.
The oversimplified assumption of uniformity along poloidal direction encounters difficulty as the currents in active antennas and their box frames break the poloidal symmetry.
Therefore, the analysis based on phase velocity estimations have inherit difficulty in capturing the poloidal variations of RF-induced electric fields and plasma potentials.
On the other hand, although not presented in this study, the TDE techniques are capable of showing the 2D distribution of RF-induced radial electric fields and the poloidal profile of RF-induced plasma potentials, and hence would be beneficial to future investigations of the RF physics.

\subsection{Broadening of Radial Electric Fields at Higher ICRF Power}

\begin{figure}
  \includegraphics[width=3.5in]{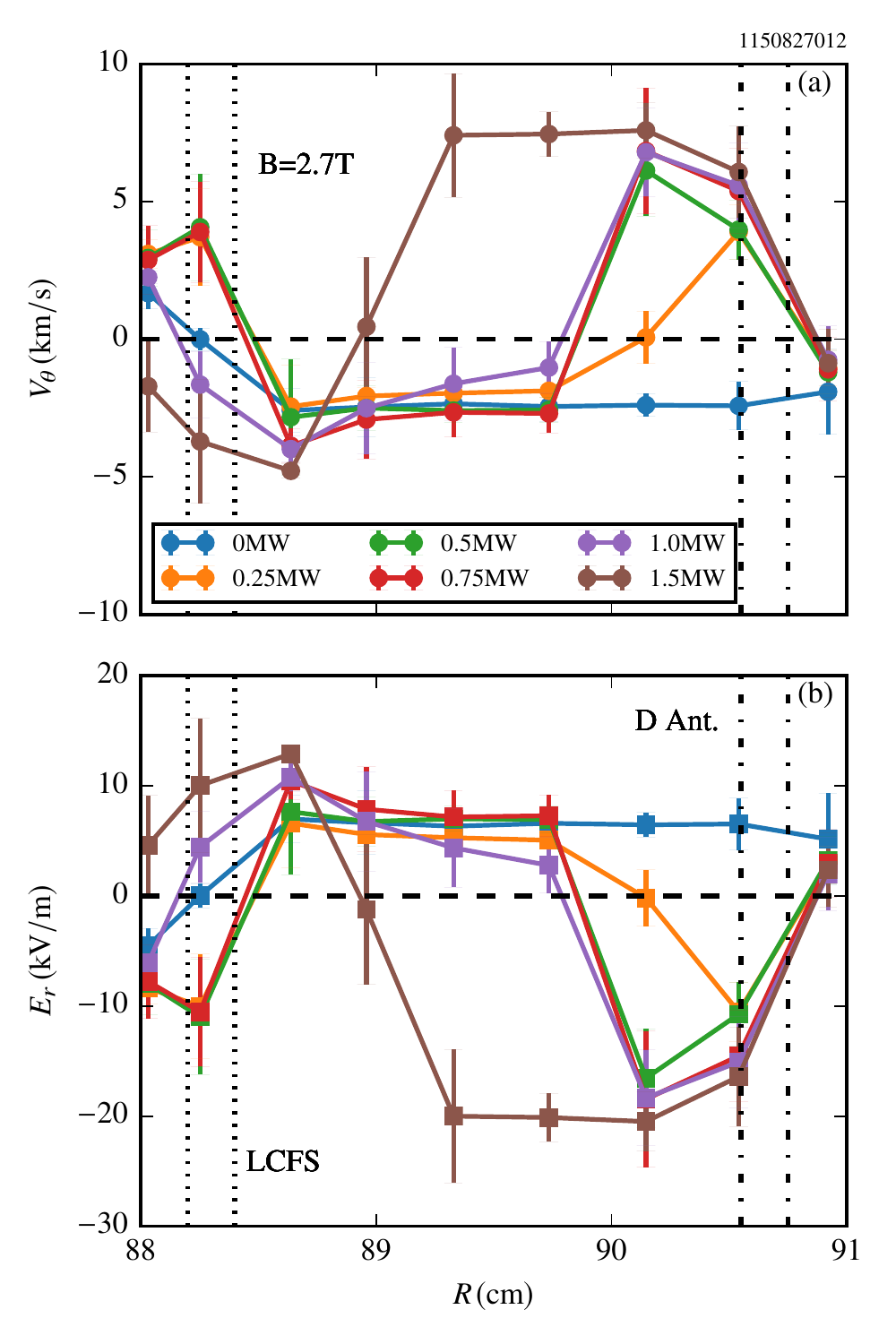}
  \caption{\label{fig:vel_Er_2T} 
    Equilibrium profiles of (a) poloidal velocity $ V_\theta(r) $ and (b) radial electric field $ E_r(r)=-V_\theta B_\varphi $, at different ICRF heating powers launched by the D antenna. The toroidal magnetic field is $ B_\varphi = 2.7\,\mathrm{T} $. Black dotted lines indicate the position of the LCFS; black dash-dot lines indicate the position of the D antenna. The RF-induced $ E_r $ field increases as RF power is raised.}
\end{figure}

The radial profiles of the poloidal velocity can be obtained by averaging the 2D velocity field along the poloidal direction.
Figure \ref{fig:vel_Er_2T}(a) shows the resulting equilibrium profiles of poloidal velocities, $ V_\theta(r) $, at $ B_{\varphi} = 2.7 $ T under different ICRF powers in L-mode plasmas.
The radial electric field can then be inferred from the poloidal velocity, $ E_r(r) = -V_\theta B_\varphi $ (shown in Fig. \ref{fig:vel_Er_2T}(b)), since the $ \mathbf{E\times B} $ flows is large compared to the flow induced by thermal sheath drop in the far SOL region \cite{Cziegler2012PPaCF105019,Ochoukov2014PPaCF15004}.
For the discharge at $ B_{\varphi} = 2.7 $ T, ICRF waves induce an inward radial electric field, $ E_r $, near the active antenna, while the direction of $ E_r $ near the separatrix remains outward.
This change in $ E_r $ is considered as a result of the ICRF rectified sheath.
Results also show that as ICRF power is raised, the magnitude of $ V_\theta $ and the RF-induced $ E_r $ become larger.

\begin{figure}
  \includegraphics[width=3.5in]{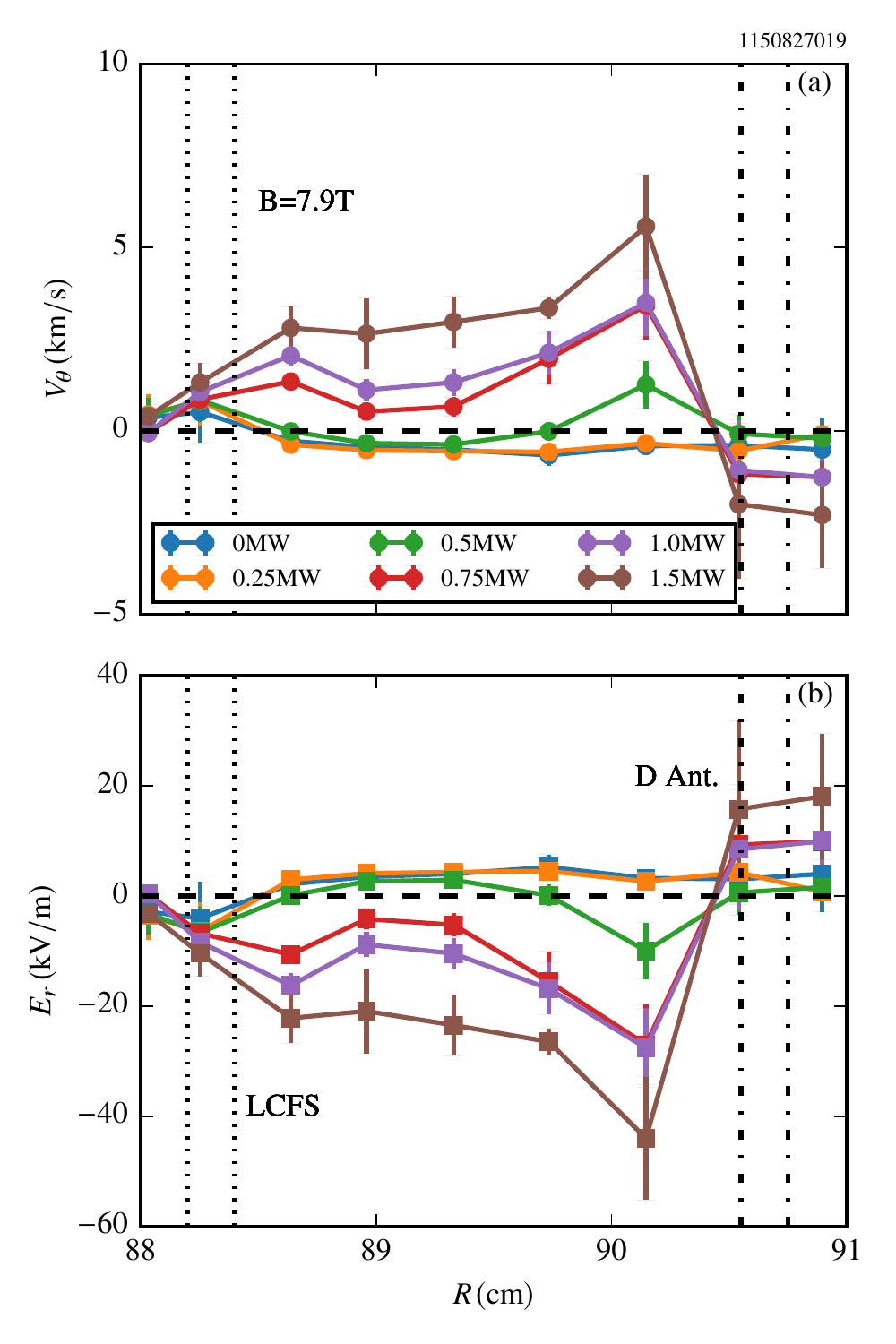}
  \caption{\label{fig:vel_Er_8T} Profiles of (a) poloidal velocity $ V_\theta(r) $ and (b) radial electric field $ E_r(r)=-V_\theta B_\varphi $, at different ICRF heating powers launched by the D antenna. The toroidal magnetic field is $ B_\varphi = 7.9\,\mathrm{T} $. Black dotted lines indicate the position of the LCFS; black dash-dot lines indicate the position of the D antenna. The RF-induced $ E_r $ field increases as RF power is raised.}
\end{figure}

Figure \ref{fig:vel_Er_8T} shows the radial profiles of poloidal velocity, $ V_\theta(r) $, and radial electric field, $ E_r(r) $, at different ICRF powers launched by the D antenna when $ B_\varphi = 7.9\,\mathrm{T} $ .
In this discharge, the radial extension of the inward $ E_r $ induced by ICRF waves is less than 1 cm when $ P_{\mathrm{ICRF}} \leq 0.5 $ MW.
But the induced $ E_r $ immediately extends to the separatrix, as $ P_{\mathrm{ICRF}} $ is raised to 0.75 MW and beyond. 

\begin{figure}
  \includegraphics{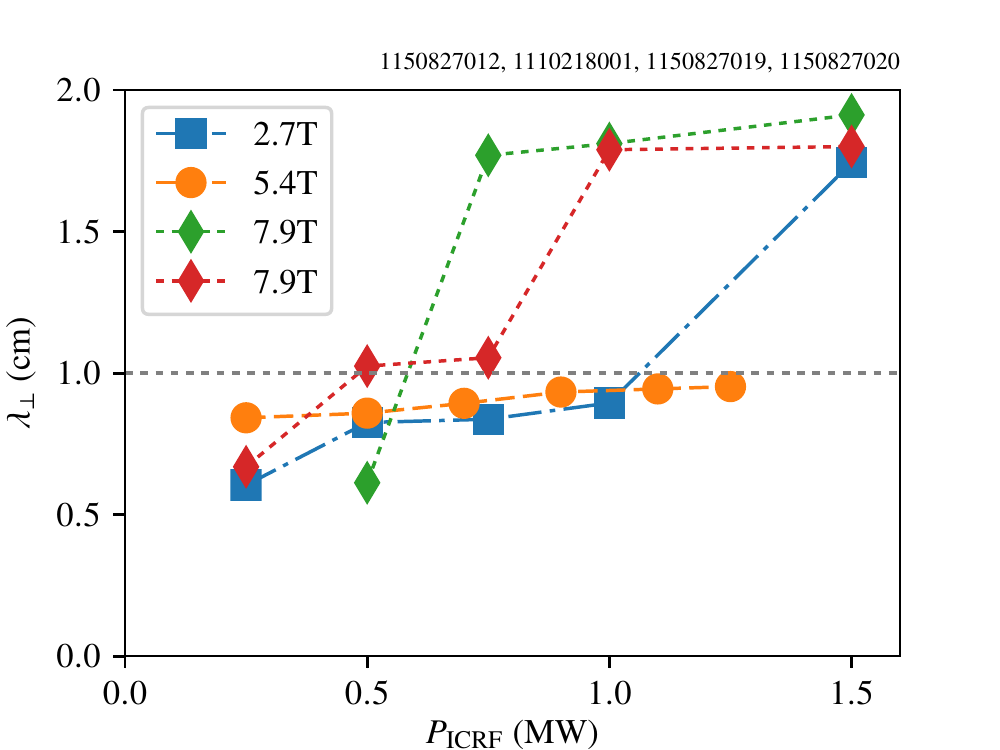}
  \caption{\label{fig:rad_scale} The radial width of the RF-induced radial electric field, $ E_r $, compared against the ICRF heating powers, at different toroidal magnetic field.}
\end{figure}

As observed from profiles plotted Fig. \ref{fig:vel_Er_2T} and \ref{fig:vel_Er_8T}, the width of inward $ E_r $ induced by ICRF waves near the antenna is $ \lambda_{\bot}\sim 1-2 \,\mathrm{cm}$ and is much larger than the local skin depth $ \delta_\mathrm{pe}=c/\omega_{\mathrm{pe}} \approx 1-3\,\mathrm{mm} $ which is the expected penetration length of the SW.
A number of theoretical models have been proposed to explain this "anomalous" penetration length of the dc potential and electric field structure \cite{Faudot2010PoP,Myra2008PRL195004,Myra2010PPaCF15003}.
A possible mechanism is due to the self-consistent exchange of the transverse RF current between neighboring flux tubes \cite{Faudot2010PoP}.
The broadening of the penetration length, according to the linear modeling, is predicted to be $ \lambda_{\bot} \sim \left(L_{\parallel} \rho_{\mathrm{ci}}/2 \right)^{1/2} $ at large RF powers.
This model yields $ \lambda_{\bot} \approx 1\,\mathrm{cm}$ at the real C-Mod conditions of $ \rho_{\mathrm{ci}}\approx 0.01 \,\mathrm{cm} $ and the connection length $ L_{\parallel} \approx 200\,\mathrm{cm} $, which is shown to be consistent with preceding GPI observations \cite{Cziegler2012PPaCF105019} at $ B_{\varphi} = 5.4\,\mathrm{T} $ (data points shown in Fig. \ref{fig:rad_scale}).
In addition, the model indicates that the radial width should be reduced at higher magnetic field ($ \lambda_{\bot} \sim \rho_\mathrm{ci}^{1/2} \sim B_{\varphi}^{-1/2} $).
However, as can be seen in Fig. \ref{fig:rad_scale}, the observed radial width of the ICRF-induced $ E_{r} $ substantially \emph{increases} from $ \lambda_{\bot} \approx 0.9 \,\mathrm{cm}$ to $ \sim 1.8 $ cm when $ B_{\varphi} $ is raised from 2.7 to 7.9 T at larger ICRF heating powers.
The radial expansion of ICRF-induced $ E_r $ at $ B_\varphi = 7.9 $ T with larger heating power covers the whole SOL region, i.e. $ \lambda_{\perp} \approx L_\mathrm{SOL} $, which corresponds to a significantly increased plasma potential, $ \phi_{\mathrm{ind}} $, near the active antenna.
This large plasma potential can accelerate ions along the magnetic field and enhance the sputtering on PFCs.

\section{ICRF-induced Plasma Potentials}

Since the ICRF-induced plasma potential arises from the rectified parallel electric field from slow waves, it is predicted to be proportional to the square root of the ICRF power \cite{Faudot2010PoP},
\[ \phi_{\mathrm{ind}} \sim P_{\mathrm{ICRF}}^{1/2} .\]
This scaling has been examined in previous power scan experiments \cite{Cziegler2012PPaCF105019,Ochoukov2013JoNM875,Ochoukov2014PPaCF15004,Wukitch2013PoP}.
This plasma potential can also be estimated by integrating $E_r$ determined by the GPI measurement.
The inferred radial electric field near the active ICRF antenna consists of two components: (a) the rectified RF field from the ICRF antenna; (b) the background electric field due to the thermal sheath drop, i.e. $ E_{r0} = - 3 \partial_{r} T_{e} / e $, which can be estimated from the ohmic plasmas and is typically small compared to the RF-enhanced part \cite{Cziegler2012PPaCF105019}.
The peak value of the induced potential can be evaluated as 
\begin{equation} 
\phi_{\mathrm{ind}} = - \int_{r_{\mathrm{sep}}}^{r_{\mathrm{ant}}} E_{r,\mathrm{ind}} \,\mathrm{d}r = B_{\varphi} \int V_{\theta,\mathrm{ind}} \,\mathrm{d}r.
\label{eq:phi}
\end{equation}
In this experiment, different heating scenarios have been utilized to study other effects on the rectified sheath potential, such as the toroidal field strength, the impurity seeding, antenna geometry and current strap phasing.

\subsection{Dependence on Toroidal Field Strength}

Recent experiments suggest that large convective cells induced by RF waves may play an important role in enhanced ICRF impurity sources and contamination \cite{Wukitch2013PoP,Cziegler2012PPaCF105019}.
The cell strength scales $ B_{\varphi}^{-1/2} $ \cite{Faudot2010PoP} and thus we expect the convective cell and peak potential to decrease with higher toroidal field strength.
To probe the convective cell strength and induced plasma potentials, we seek to characterize the dependence of the SOL turbulence poloidal velocity on toroidal field using GPI imaging data and TDE techniques.

\begin{figure}
  \includegraphics{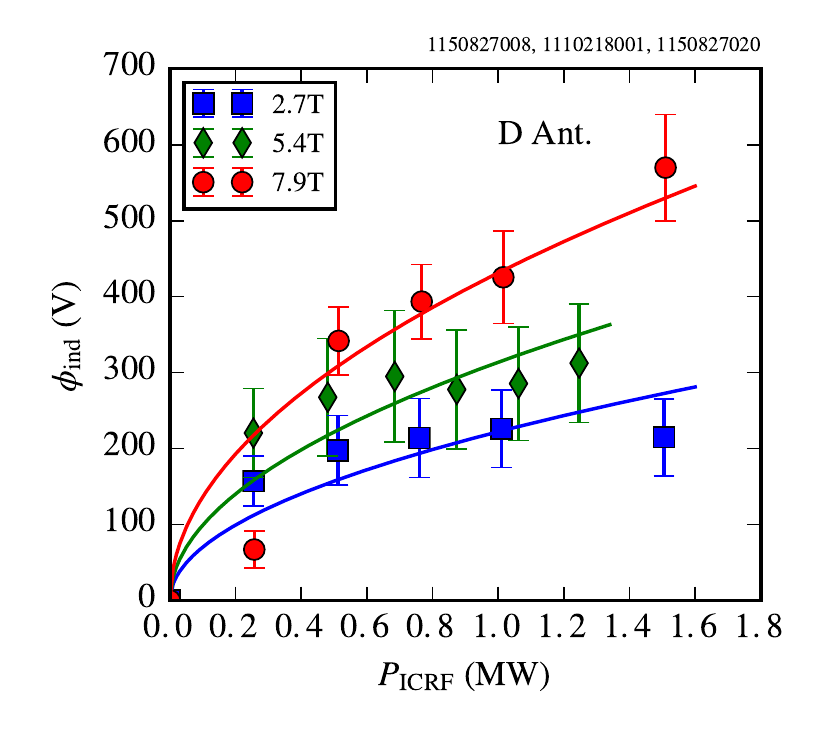}
  \caption{\label{fig:pot_TA_B} 
    Peak values of RF-induced plasma potential as a function of ICRF heating power at three different toroidal magnetic fields. Markers represent experimental estimates and solid lines represent their best fits in the form of $ P_{\mathrm{ICRF}}^{1/2} $. The error bars indicate two standard errors of the mean values.}
\end{figure}

Estimated peak values of the ICRF-enhanced plasma potentials that are induced by the D antenna are plotted in Fig. \ref{fig:pot_TA_B}, as a function of the ICRF powers, at three different toroidal magnetic fields, 
$ B_{\varphi} = 2.7,\, 5.4\, \mathrm{and}\, 7.9\,\mathrm{T}$.
To achieve a similar safety factor and plasma shape, the plasma current $ I_{p} $ was scaled with $ B_{\varphi} $.
The induced potentials $ \phi_{\mathrm{ind}} $ scaled approximately as $ P_{\mathrm{ICRF}}^{1/2} $, which is in agreement with the model of rectified sheath potentials.
As $ B_{\varphi} $ was increased from 2.7 to 7.9~T, $ \phi_{\mathrm{ind}} $ increased substantially.
This could be partly attributed to the significant radial expansion of $ E_r $ at higher $ B_\varphi $.
In addition, the measured $V_\theta$ did \emph{not} scale down as the $B_\varphi$ was raised, which could also contribute to the large plasma potential.

In all cases the induced potentials were much larger than the background thermal sheath potential ($ 3T_{e} \sim 30\,\mathrm{eV} $).
In particular, for $ B_{\varphi} = 7.9\,\mathrm{T} $ and $ P_{\mathrm{ICRF}} = 1.5\,\mathrm{MW} $, the maximum of the induced potential was about 600 V.
The large induced potential can enhance the sputtering yield on the PFCs and thus increase the background impurity content considerably \cite{Wukitch2009JoNM951,Ochoukov2013JoNM875}.
This result presents a challenge to the control of the plasma-material interaction in high-field fusion devices, as well as the concepts of high-field-side launch of RF waves, which are expected to yield a better heating efficiency of thermal ions without generating energetic minority ion tails \cite{Wallace2015ACP30017}.

Complicating the interpretation of the RF enhanced potential dependence on $ B_\varphi $ are the differences in ICRF absorption scenarios and edge conditions.  In C-Mod, different $ B_\varphi $ correspond to different ICRF absorption mechanisms in these experiments. 
For 2.7 T, second harmonic H minority heating is utilized whereas fundamental minority H and $ ^3\mathrm{He} $ heating are used at 5.4 T and 7.9 T, respectively.
In C-Mod, the RF power absorption effectiveness in the plasma core is highest for H minority heating whereas both the second harmonic H minority and fundamental minority $^3\mathrm{He} $ heating are both substantially weaker.
Within this data set, the lowest and highest RF enhanced potentials correspond with the weaker central absorption indicating that dependence on $ B_\varphi $ is unlikely to be a result of differences between core absorption scenarios.
The normalized density, ratio of density to the Greenwald density $ \bar{n}/n_G $, for the three discharges shown is 0.4-0.44, 0.23-0.28, and 0.21-0.23 for the 2.7 T, 5.4 T, and 7.9 T discharges respectively.
The RF enhanced potentials are lowest for the largest normalized density and highest for the smallest normalized density.
This suggests the plasma edge conditions play a role as well as the RF fields.
The understanding, however, between strength of RF absorption, plasma edge conditions and induced plasma potential is incomplete and requires further studies.

\subsection{Effects of Impurity Seeding}

\begin{figure}
  \includegraphics{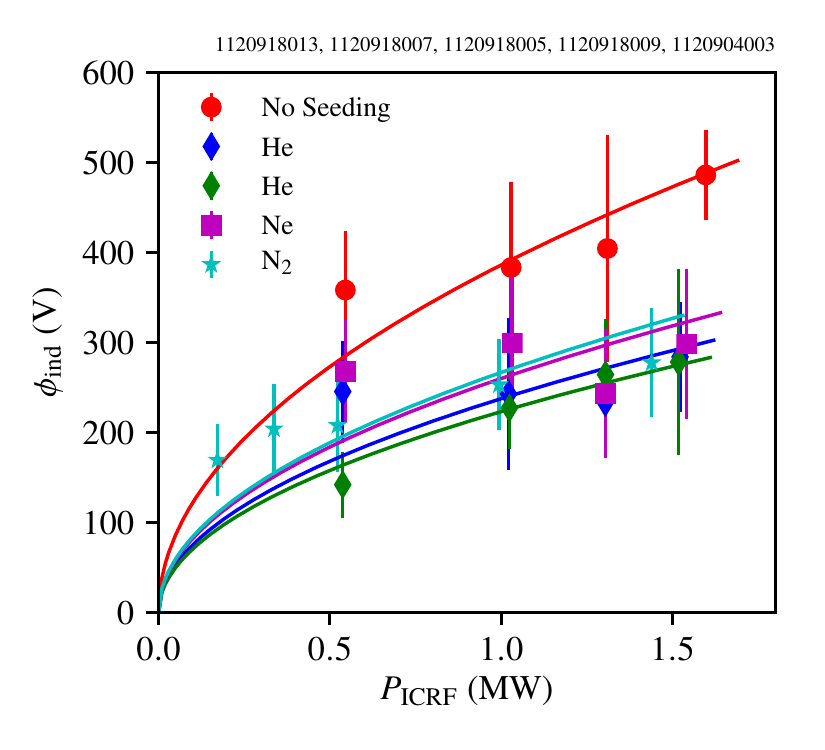}
  \caption{\label{fig:pot_impurity}
    Peak values of RF-induced plasma potential as a function of ICRF heating power of D antenna with different species of low-Z impurity seeding. $ B_{\varphi}= 5.4\,\mathrm{T} $ and $ I_{p} = 0.8\,\mathrm{MA}$. Markers represent experimental estimates and solid lines represent their best fits in the form of $ P_{\mathrm{ICRF}}^{1/2} $.}
\end{figure}

The performance of ICRF-heated plasmas was found to be improved with impurity seeding in previous experiments \cite{Greenwald2014PoP}.
Contributing to this effect is the reduction of ICRF-specific impurities particularly core Mo concentration and reduced core impurity radiation.
To investigate the response of the RF-induced plasma potential to the impurity seeding, low-Z gases (helium, nitrogen and neon) were puffed near the antenna.
The seeding gases were puffed with a pressure of 2 psi and a pulse duration of 200 ms at 50 V on the piezo.
Figure \ref{fig:pot_impurity} shows the estimated RF sheath potentials induced by the TA D antenna with different kinds of low-Z impurities.
The peak values of RF-induced plasma potentials decreased by about 30\% when impurity gases are injected.
Moreover, with impurity seeding the response of the RF-induced potential to the ICRF power deviates from the $ P_{\mathrm{ICRF}}^{1/2} $ scaling considerably.
A possible cause of the reduction in RF sheath potential is that high neutral contents may substantially increase the collisionality/resistivity and damp the plasma flows in the SOL region.
High collisionality/resistivity will limit the transverse current between neighboring flux tubes, and thus inhibit expansion of the radial electric field and the plasma potential induced by the ICRF waves.

Injecting impurity gases usually increases the plasma density in the SOL region.
The increase in the edge density, as predicted by the SW rectification model, will increase the RF sheath potential, as long as the edge density is greater than a threshold (about $ 10^{16}\,\mathrm{m^{-3}} $ in C-Mod) \cite{Myra2008PRL195004,Myra2010PPaCF15003}.
The threshold has been observed by the probe measurements in C-Mod \cite{Ochoukov2014PPaCF15004}, however, the GPI observations in the impurity seeding experiments do not agree with the model.
Clearly, further work is needed to understand the effects of the impurity seeding on the RF sheath.

\subsection{Field-Aligned Antenna vs. Toroidally-Aligned Antenna}

\begin{figure}
  \includegraphics{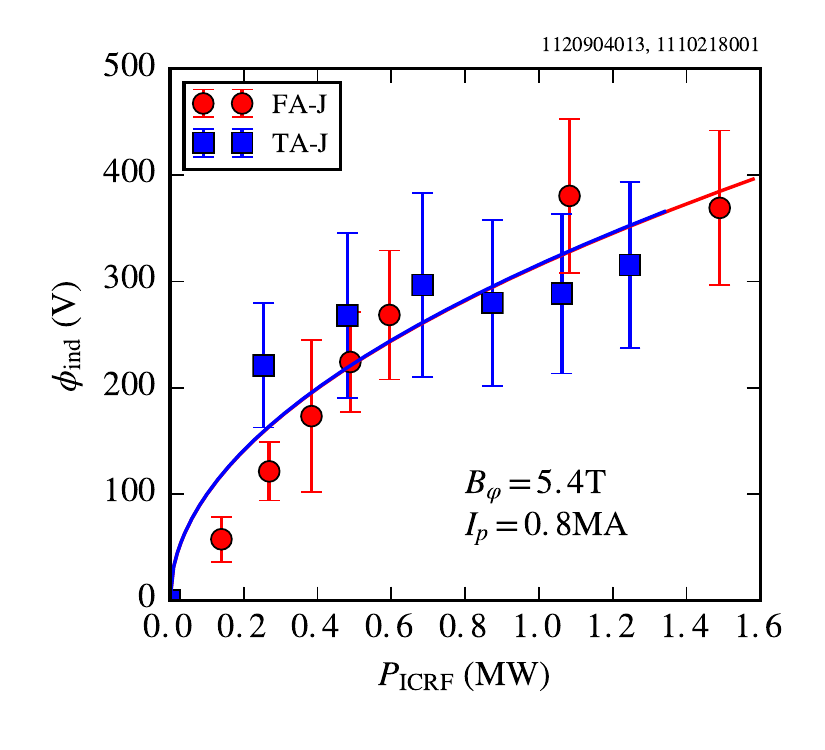}
  \caption{\label{fig:pot_FA_TA} 
    Peak values of the RF-induced sheath potential as a function of ICRF heating power by FA and TA antenna. Markers represent experimental estimates and solid lines represent their best fits in the form of $ P_{\mathrm{ICRF}}^{1/2} $.}
\end{figure}

To reduce the RF sheath potential and the impurity content, C-Mod has recently implemented the design of the field-aligned (FA) antenna \cite{Garrett2012FEaD1570}.
The FA antenna is distinguished from the conventional toroidally-aligned (TA) antenna by current straps and an antenna box structure that are perpendicular to the total magnetic field.
Such a symmetrical design, as predicted by 3D antenna modeling, are expected to minimize the integrated $ E_{\parallel} $ (electric field
along a magnetic field line), and lead to a reduction of the RF-induced plasma potential \cite{Garrett2012FEaD1570,Wukitch2013PoP}.
The RF-induced plasma potentials calculated using TDE techniques are shown in Fig.~\ref{fig:pot_FA_TA}, which is consistent with previous analysis based poloidal phase velocity estimations \cite{Wukitch2013PoP}.
Although the reduction of the impurity level has been observed in the plasmas heated by FA antenna \cite{Wukitch2013PoP,Greenwald2014PoP}, the potential induced by FA antenna was similar to that by TA antenna in present study.
The observations indicate that, although the reduction in the integrated $ E_\parallel $ can reduce the release of ICRF-specific impurities, the local RF-$ E_{\parallel} $ and RF sheath potential may not decrease.

On the other hand, we note that recent experiments in ASDEX Upgrade also show a significant reduction of the impurity tungsten (W) released from ICRF antennas when the new 3-strap antenna (toroidally-aligned) is used \cite{Bobkov2016NF84001}.
Particularly, a minimum W content in core plasma was achieved as the power ratio between the central strap and the outer straps was varied from $ \frac{1}{1} $ to $ \frac{3}{1} $.
This narrow range of the power ratio corresponds to a cancellation of RF image 
currents and therefore minimizations of both local RF-$ E_\parallel $ fields and RF sheath potentials on the side limiters \cite{Bobkov2017PPaCF14022}.
These findings imply that the significant plasma potentials induced by FA antennas may result from the image currents on the side limiters and antenna box structures.

\subsection{Monopole Phasing vs. Dipole Phasing}

\begin{figure}
  \includegraphics{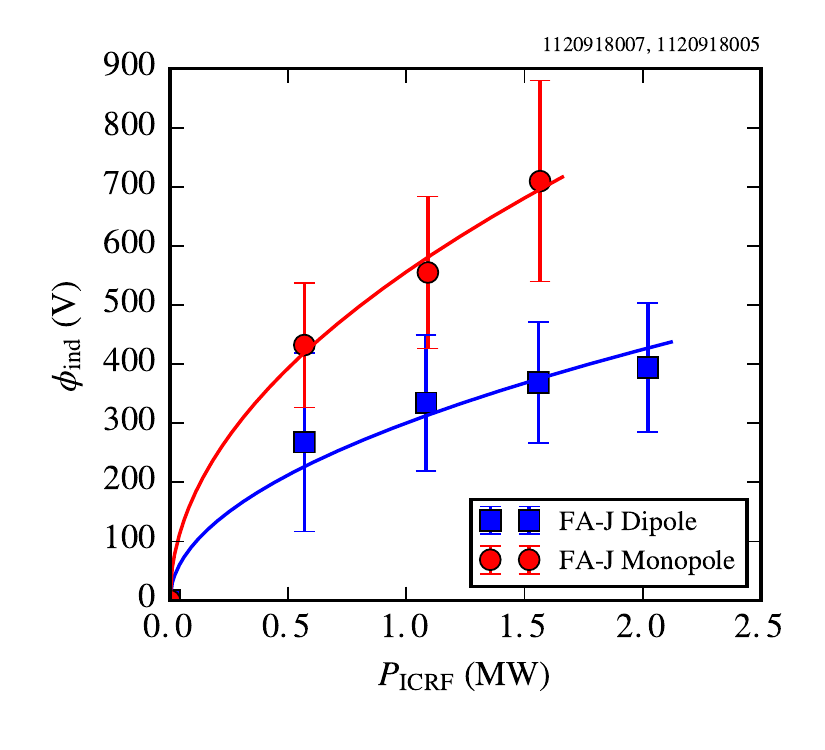}
  \caption{\label{fig:pot_di_mono} 
    Peak values of RF-induced sheath potential as a function of ICRF heating power in dipole and monopole phasing. $ B_{\varphi}= 5.4\,\mathrm{T} $ and $ I_{p} = 0.8\,\mathrm{MA}$. Markers represent experimental estimates and solid lines represent their best fits in the form of $ P_{\mathrm{ICRF}}^{1/2} $.}
\end{figure}

According to 3D antenna modeling, the reduction in integrated $ E_{\parallel} $ depends on the relative phases of the current straps \cite{Wukitch2013PoP}.
For the FA antenna, the monopole phasing is predicted to exhibit the lowest integrated $ E_{\parallel} $.
The TDE techniques have also been used to estimated the RF-induced plasma potentials, and show consistent results with previous analysis \cite{Wukitch2013PoP}.
As demonstrated in Fig.~\ref{fig:pot_di_mono}, the measured plasma potential induced in the monopole phasing was significantly larger than in the dipole phasing.
Consistent with the behavior of RF-enhanced plasma potentials, the local impurity source from the antenna in the monopole phasing is also found to be larger than that in the dipole phasing \cite{Wukitch2013PoP}.
Again, this substantially higher value of RF sheath potential might arise from the RF image currents on the FA antenna box structures.
The lack of poloidal symmetry in monopole phasing antenna might also lead to larger image currents and RF sheath potentials.

\section{Summary}

In this study, the ICRF-induced plasma potentials and radial electric fields
have been investigated using gas-puffing imaging techniques in the Alcator C-Mod tokamak.
The equilibrium plasma flows that advect turbulent structures in SOL region has been calculated using TDE techniques.
The results of radial profiles from TDE analysis are in agreement with previous analysis based on poloidal phase velocity estimations.
The TDE techniques are able to show 2D distributions of RF-induced $ E_{r} $ and $ \phi_{\mathrm{ind}} $, and will be beneficial for future studies of RF physics.

The large convective cells induced by the RF sheath electric field are observed in the vicinity of the active ICRF antenna.
The radial width of this field is $ 1-2\,\mathrm{cm} $ and \emph{increases} as the toroidal magnetic field is raised.
In most discharges, the peak values of the ICRF-induced potential, $ \phi_{\mathrm{ind}} $, scale as $ P_{\mathrm{ICRF}}^{1/2} $ and are proportional to the strength of the toroidal magnetic field.
In particular, $ \phi_{\mathrm{ind}} \approx 600 \,\si{\volt}$ is observed at $ B_\varphi = 7.9\,\si{\tesla} $.
However, $ \phi_{\mathrm{ind}} $ was decreased by sufficiently strong low-Z impurity seeding, and the potentials began to deviate from the $ P_{\mathrm{ICRF}}^{1/2} $ scaling.
The sheath potential induced by the FA antenna was similar to that by TA antenna in disagreement with the modeling which indicate that the FA antenna should produce lower integrated $ E_{\parallel} $.
In addition, while monopole FA antenna modeling predicts lower integrated $ E_{\parallel} $ than dipole phasing modeling, it induced substantially higher sheath potential in our experiments.
These discrepancies between the observations and the modeling may be attributed to the RF image currents on the FA antenna box structures, but further work is needed to test these conjectures.	

The large $ \phi_{\mathrm{ind}} $ at high toroidal magnetic field is expected to enhance the sputtering yields on PFCs adjacent to the active antenna.
Therefore, new designs of ICRF antenna should try to minimize the local RF-$ E_\parallel $ fields and related plasma-material interactions close to the antenna.

Finally, the measurements of the radial broadening of the RF sheath structure at different magnetic field do not support the scaling $ \lambda_{\bot} \sim \left(L_{\parallel} \rho_{\mathrm{ci}}/2 \right)^{1/2} $ predicted by theoretical modeling.
Additional theory and experimental work is required to clarify the underlying physics.

\section*{Acknowledgment}

The authors greatly appreciate the effort and support of the entire Alcator C-Mod team in performing these experiments.
This work is supported by the U.S. DoE, Office of Science, Office of Fusion Energy Sciences, User Facility Alcator C-Mod under DE-FC02-99ER54512 and DE-SC 0010720.

\bibliography{rf_sheath_physics}	

\end{document}